\documentclass[12pt,preprint]{aastex}

\usepackage{rotating}

\slugcomment{accepted by AJ}

\shorttitle{An occulting pair in ANGST}
\shortauthors{Holwerda et al.}

\begin{document}

\title{An extended dust disk in a spiral galaxy;\\ An occulting galaxy pair in ANGST }
\author{B. W. Holwerda\altaffilmark{1}, W. C. Keel\altaffilmark{2}, B. Williams\altaffilmark{3}, 
J. J. Dalcanton\altaffilmark{3}, R. S. de Jong\altaffilmark{1}}
\email{holwerda@stsci.edu}

\altaffiltext{1}{Space Telescope Science Institute, Baltimore, MD 21218, USA}
\altaffiltext{2}{Department of Physics \& Astronomy, 206 Gallalee Hall, 514 University Blvd., University of Alabama, 
Tuscaloosa, AL 35487-0324, USA}
\altaffiltext{3}{Physics-Astronomy Bldg., 3910 15th Ave NE, Room C319, Seattle WA 98195-0002 , USA}

\begin{abstract}
We present an analysis of an occulting galaxy pair, serendipitously discovered in ACS Nearby Galaxy Survey Treasury (ANGST) observations of NGC 253  taken with Hubble Space Telescope's Advanced Camera for Survey in $F475W$, $F606W$ and $F814W$ ($SDSS-g$, broad $V$ and $I$). The foreground disk system (at $z \le 0.06$) shows a dusty disk much more extended than the starlight, with spiral lanes seen in extinction out to 1.5 $R_{25}$, approximately six half-light radii. 
This pair is the first where extinction can be mapped reliably out to this distance from the center.
The spiral arms of the extended dust disk show typical extinction values of $A_{F475W} \sim 0.25$, $A_{F606W} \sim 0.25$, and $A_{F814W} \sim 0.15$. The extinction law inferred from these measures is similar to the local Milky Way one, and we show that the smoothing effects of sampling at limited spatial resolution ($<57$ pc, in these data) flattens the observed function through mixing of regions with different extinction. This galaxy illustrates the diversity of dust distributions in spirals, and the limitations of adopting a single dust model for optically similar galaxies.
The ideal geometry of this pair of overlapping galaxies and the high sampling of HST data make this dataset ideal to analyze this pair with three separate approaches to overlapping galaxies: (A) a combined fit, rotating copies of both galaxies, (B) a simple flip of the background image and (C) an estimate of the original fluxes for the individual galaxies based on reconstructions of  their proper isophotes. We conclude that in the case of high quality data such as these, isophotal models are to be preferred.

\end{abstract}




 
%







\section{Introduction}

Dust absorbs starlight and re-emits it in the infrared, producing profound effects on the observed light of galaxies.
The degree to which dust extinction affects observations of their stellar disks was a source of contention in the 
early 1990's \citep{Disney89, Valentijn90} but a consensus was soon reached; disks are semi-transparent within much of the 
optical radius, but with more opaque spiral arms \citep{Cardiff94}. However, the radial extent, distribution and 
evolution of the dust in spirals at large remain poorly known. While progress can be made using mid- and far-infrared emission from dust, such maps depends strongly on the availability of local dust heating sources, rather than solely the total amount of dust. As an alternative, one can map the extinction caused by dust directly, provided one can identify a suitable background source. The challenge is therefore to find a sufficiently smooth and known background light source with which to identify and characterize dust extinction in a foreground galaxy. 

With a method proposed by \cite{kw92}, one can estimate  the extinction, using cases where a foreground spiral is observed partially in front of a background galaxy. The few suitable galaxy pairs known at the time were studied with ground-based imaging \citep{Andredakis92, Berlind97, kw99a, kw00a}, spectroscopy \citep{kw00b}, and subsequent Hubble Space Telescope (HST) observations \citep{kw01a, kw01b, Elmegreen01}. The fact that spiral disks are semi-transparent, with more opaque spiral arms, is in part based on these studies and confirmed with distant galaxy 
counts through the disks of spirals \citep{Holwerda05a, Holwerda05b, Holwerda05c, Holwerda05d, Holwerda05e, Holwerda07a, Holwerda07b}. The dust appears to be distributed in a fractal pattern and the extinction law in normal spirals is close to the Milky Way law, provided the extinction is measured on scales of 60 pc or less \citep{kw01a}. 

The occulting galaxy technique can be used to address outstanding questions about the distribution of dust-rich 
interstellar medium in spiral galaxies. First, what is the radial extent of dust in spirals? The interstellar matter (ISM) 
extends to large radii in the form of atomic hydrogen (HI) but thus far there is little direct evidence of dust beyond 
the optical disk (de Vaucouleurs' $R_{25}$).  
Some indications have been found however, partly through studies of the thermal emission by dust. For example, \cite{Nelson98} 
stacked InfraRed Astronomical Satellite (IRAS) profiles of hundreds of galaxies and found dust emission beyond $R_{25}$. \cite{Cuillandre01} reported 
reddening in the outer parts of M31's HI disk. \cite{Gordon03} find that there is a second ring structure just outside 
$R_{25}$ in M31, \cite{Regan04} finds a similar ring in NGC 7321 and \cite{Hinz06} present the best evidence to date in {\em Spitzer} data for a radially extended cold dust component in the dwarf UGC 10445. While the existing occulting-galaxy studies typically find dust
scale lengths comparable to the optical starlight, the sample is too sparse to tell much about how the dust distribution might vary within the disk-galaxy population.

There is also indirect evidence of dust at large radii from observations of extended star formation in spiral disks. 
The existence of excess-ultraviolet (XUV) disks detected with GALEX \citep{Gil-de-Paz05, Thilker05b, Thilker07} suggests 
the presence of a cold molecular-gas component at large radii, confined to spiral arms. This molecular component is also 
likely to host dust --suggesting that there may well be extinction at these large radii-- at least in spiral structures.

The second question that can be addressed in occulting pairs is: How is the dust distributed spatially? 
For the inner parts of spiral disks the distribution of the dust-rich ISM is well traced by the emission observed with 
{\em Spitzer} \citep{Dale05,Dale07,Draine07}. However, studies of warm dust leaves the structure of the cold ISM traced by dust 
unexplored over much of the HI disk. In contrast, backlit dust can be detected independent of its temperature.
In these regions, extinction studies that use occulting pairs, are the a reliable probe of the dust distribution. 
The size distribution of dust extinction can then be compared to the sizes of giant molecular clouds 
\citep{Rosolowsky05}.


The final outstanding question is how the opacity of spiral disks changes over time. There was 
more star formation at $z \approx 1$ by an order of magnitude  \citep{Madau98,Steidel99, Giavalisco04, Thompson06, 
Hopkins06}. The resulting increase in the gas phase metallicity should result in more dust and consequently, in an 
increase in the average disk opacity with time. Some models suggest a peak in dust content and opacity at
intermediate redshifts, as some galaxies exhaust their interstellar gas \citep{Calzetti99}.  
\cite{Holwerda07c} sought to address this using ideal (spiral/elliptical) occulting pairs from the Sloan Digital Sky 
Survey. They estimate the disk radial opacity as a function of redshift to a redshift of z=0.2 and find no evolution. 
With HST, occulting pairs can be studied up to a redshift of z $\sim$ 1, provided a sufficient number of pairs can be 
found. Local pairs such as the one presented here, serve as the local reference frame to interpret the more distant 
pairs and as a test of different methods for analyzing the extinction.

In this paper, we report on an occulting pair with nearly ideal geometry. The background galaxy ($z=0.06$) is a 
spiral with a prominent, regular and smooth bulge, which is partially occulted by a foreground spiral. The pair 
was serendipitously discovered in HST images from the ACS Nearby Galaxy Survey Treasury (ANGST) survey. 
We explore the radial and spatial distribution and the inferred extinction law of the dust in a smaller foreground 
spiral, which displays an extent of dust hitherto unseen in backlit galaxies. 
In \S \ref{s:data} we present the data; in \S \ref{s:mod} the analysis and different approaches; and in \S \ref{s:map}, 
the optical depth estimates. We discuss the  results and approaches in \S \ref{s:dis} and list our conclusions 
in \S \ref{s:con}.

\section{\label{s:data}Data}

\begin{figure}[h]
\begin{center}
  \includegraphics[width=0.8\textwidth]{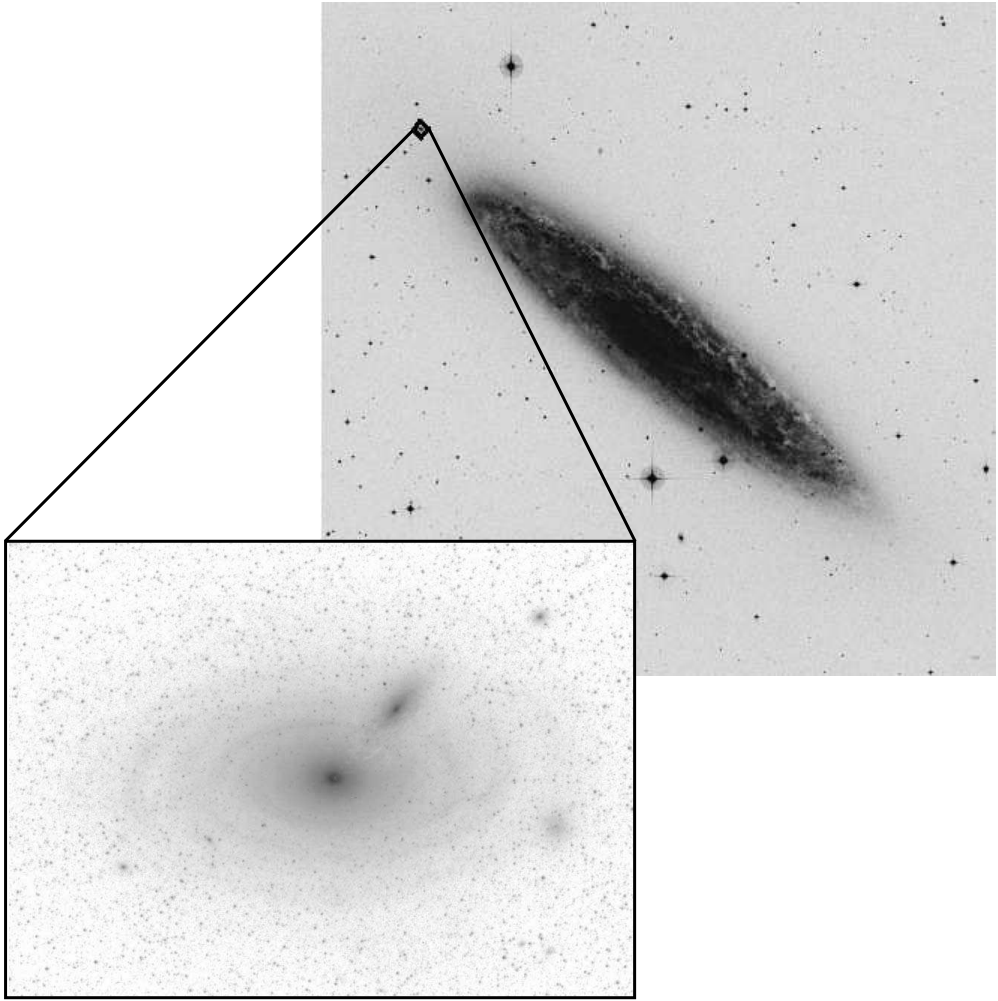}
\caption{The position of our occulting pair with respect to NGC 253. The Digitized Sky Survey image of NGC 253 retrieved from NED. The blow-up is a grayscale version of the color image, Figure \ref{f:rgb}. }
\label{f:n253}
\end{center}
\end{figure}

The HST/ACS data were obtained for the ACS Nearby Galaxy Survey Treasury \citep[ANGST,][]{angst}, as part of 
their observations of NGC 253 to characterize stellar populations as a function of radius (Figure \ref{f:n253}). Data 
are in three filters: $F475W$, $F606W$ and $F814W$, corresponding approximately to $SDSS-g$, broad $V$ and 
$I$. Exposure times are  2256, 2283, and 2253 seconds respectively. Figures \ref{f:n253} and \ref{f:rgb} show a 
grayscale and color composite of the three filters. The image shows clear dust structures associated with the 
foreground spiral, extending to (and slightly beyond) the center of the bulge of the background galaxy. 
Figure \ref{f:map} shows the $F475W$ band grayscale image in which the dust lanes of the foreground galaxy 
are neatly visible against the background galaxy's bulge.

The background galaxy was catalogued as 2MASXJ00482185-2507365 by 2MASS. \cite{Ratcliffe98} give its redshift 
as $z=0.06$. At this distance a single ACS pixel ($0\farcs05$) corresponds to a physical scale of 57 pc., assuming 
$h_0 = 70 km/s / Mpc.$ The redshift of the foreground companion spiral is not known, so we take $z=0.06$ as an 
upper limit when computing physical scales within the foreground system.

\section{\label{s:mod}Analysis}

\begin{figure}[h]
\begin{center}
  \includegraphics[width=0.8\textwidth]{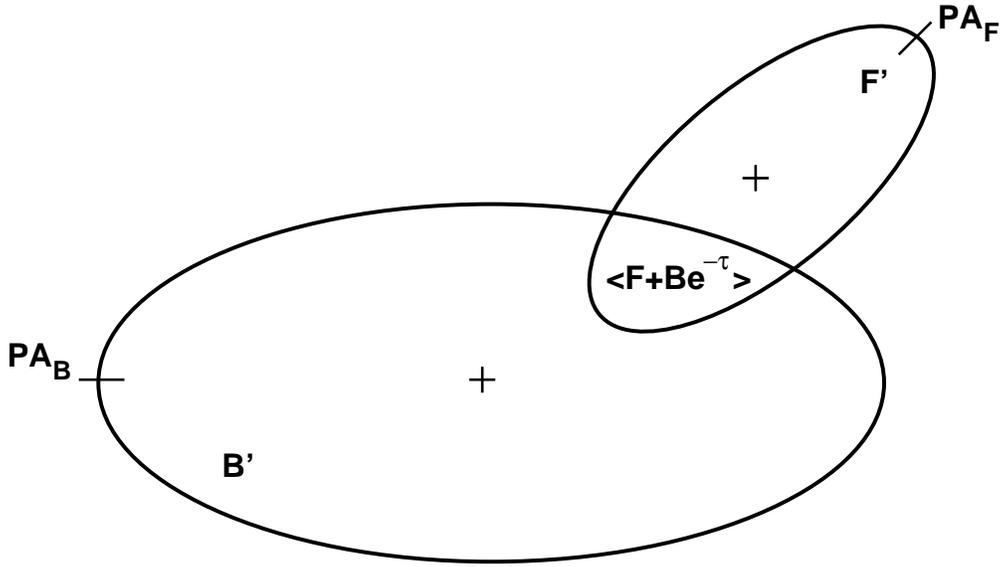}
\caption{A cartoon of the occulting galaxy method. There are three flux measurements in the image: the overlap region ($F+Be^{-\tau}$), and the non-overlap parts of the background galaxy ($B'$) and foreground galaxy ($F'$). The optical depth of the foreground galaxy in the overlap region is estimated from these three observed fluxes, assuming symmetry for both galaxies ($B \approx B'$, $F \approx F$), from: $e^{-\tau'} = {<F+Be^{-\tau}> -F' \over B' }$. If the background galaxy is much brighter than the foreground disk ($B >> F$), the optical depth estimate simplifies to $e^{-\tau'} = {<F+Be^{-\tau}>  \over B' }$, as used in the case of Method B. In the case of Method A, the free parameters for the fit are the position angles of the rotated foreground and background galaxy ($PA_F, ~ PA_B \sim 180$) and the centers of both galaxies. }
\label{f:model}
\end{center}
\end{figure}

To map the extinction in the foreground galaxy, we assume that the foreground and background galaxies are rotationally 
symmetric around their respective centers. This assumption holds well for ellipticals and generally well for grand-design spirals
(with rotational symmetry rather than axisymmetry). 
Figure \ref{f:model} shows a sketch of the method, where $F$ and $B$ indicate the unobscured flux due to the foreground 
and background galaxy, respectively. The flux in the overlap region comes from both galaxies, with the background galaxy 
dimmed by the opacity of the foreground disk with an optical depth $\tau$: $<F ~+~ B e^{-\tau}>$. Assuming appropriate symmetry, we can estimate the 
local optical depth using the flux from the overlap region and the corresponding point-symmetric sections in the 
foreground ($F'$) and background ($B'$) galaxy:

\begin{equation}
\label{eq1}
e^{-\tau'} = {<F ~+~ B e^{-\tau}> -  F' \over B'}.
\end{equation}

When the contribution from the foreground galaxy becomes negligible compared with the background galaxy ($B ~ >> ~ F$), 
this can be approximated by \citep{kw00a}:

\begin{equation}
\label{eq2}
e^{-\tau'} = {B e^{-\tau} \over B'}.
\end{equation}

The key step in this analysis is clearly estimating F and B at each location in the overlap region. 
Using this formalism, we consider three approaches, whose applicability depends on the geometry of the
system and the data quality. 
The first approach (method A) is to rotate both background and foreground galaxies $180^\circ$ and then subtract 
both from the original image, minimizing the residual image. The optical depth is then obtained separately from 
equation \ref{eq1} \citep[see also][]{Holwerda07c}.
The second approach (method B) assumes the contribution of the foreground galaxy is negligible. The background 
galaxy is rotated by $180^{\circ}$, and the optical depth is obtained from the ratio of the original and rotated 
image in equation \ref{eq2}. This method provides a lower limit to the opacity, because any foreground
light will masquerade as background light, reducing the inferred extinction.
The third option (method C) is to model the background galaxy with an appropriate isophotal model 
(e.g., with a S\'ersic profile, or via elliptical fits to a set of isophotes), and then to estimate the galaxies' 
contributions ($B'$ and $F'$) from the model before estimating the optical depth from equation \ref{eq1} or \ref{eq2}. 

The appropriate approach to analyze the dust structure of the foreground galaxy depends on the geometry of the 
pair. In the pair considered in this paper, we can use any of the three approaches because the background bulge 
completely dominates the flux in the outer overlap region of the foreground spiral, and there are large 
unobscured regions of both galaxies. Thus, we use these data to compare the performance of these three
approaches.
Using each of these, we will discuss the spatial extent, distribution of the extinction, and the inferred extinction 
law. For this pair, we expect methods B and C to give more reliable results, given that the assumption of axisymmetry 
is not as well justified for the outer regions of the foreground galaxy.

\subsection{\label{s:moda}Method A: Fit Rotated Galaxies}

To construct the extinction map with the first approach, we applied the method described in \cite{Holwerda07c}. First, 
we ran Source Extractor \citep{se,seman} on the image to segment it into different objects. Those sections assigned to 
either galaxy are copied to be used in the fit.
The script uses the central positions of the background and foreground galaxy and their rotation angles ($PA_B, ~ PA_F$) 
as fit parameters to minimize the flux in the residual image after both rotated galaxies are subtracted from the original.
If both galaxies are perfectly symmetric and there is no dust extinction at all, the residual image should have zero flux. 
Simply rotating copies of both galaxies and minimizing the flux in the overlap region can be effectively automated, 
enabling its application to many pairs. 
We applied separate fits in all three HST filters and the resulting optical depth maps are presented in 
Figure \ref{f:map}. The best-fit rotation angles were 180.67 and 180.74 degrees.

\subsection{\label{s:modb}Method B: Flipped Background Galaxy}

For the second approach, we rotated the background galaxy $180^{\circ}$ and constructed an optical depth image 
of the overlap region from the ratio between the original and the flipped image (Figure \ref{f:map}, second row). 
No fit was performed, and the center of the background galaxy was estimated visually.
This approach is appropriate for a quick analysis, and can be done in cases where one has the luxury of a near-perfect 
geometry and a background galaxy that is both bright and smooth.

\subsection{\label{s:modc}Method C: Isophotal Models}

\begin{figure}[h]
\begin{center}
  \includegraphics[width=0.8\textwidth]{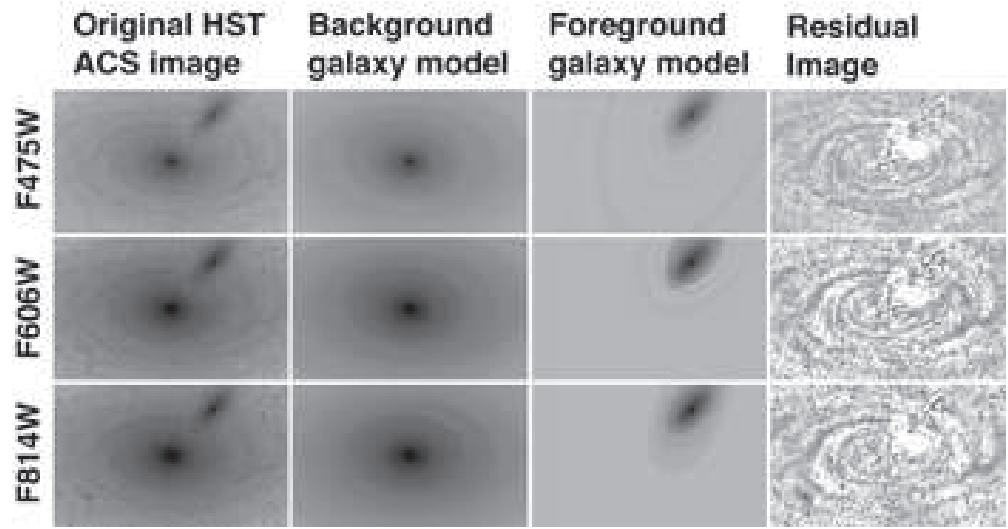}
\caption{The original image, background and foreground isophotal model and the residual image (the original image with both models subtracted) for the three HST filters, $F475W$, $F606W$, and $F814W$. Images are shown to the same logarithmic grayscale. In the residual images, the dust structures in the foreground and the spiral structure in the background galaxy stand out.}
\label{f:isomod}
\end{center}
\end{figure}

To estimate  the underlying values of $F$ and $B$, we used the {\it stsdas} {\em ellipse} and {\em bmodel} tasks 
to generate smooth models for both galaxies. First, the areas affected by the foreground galaxy were masked, and 
a set of elliptical isophotes were fit to the background galaxy. We then subtracted a noiseless realization of 
this mean profile, and modeled the foreground galaxy after masking off areas where absorption might be important, notably the overlap region and dark lanes in the background galaxy. 
The masking typically left more than 60\% of each isophote. Subsequently, we subtracted the model of the foreground 
galaxy from the data, and refit the background galaxy. If some masked regions were missed in this first iteration, 
these are masked in a second iteration. Original images, background and foreground isophotal models and the residual image are shown in Figure \ref{f:isomod} for all three HST filters. We used the isophotal model galaxies to construct an optical depth map from 
equation \ref{eq1} using the model values for $B'$ and $F'$. The resulting optical depth maps are presented in 
Figure \ref{f:map}.

The isophotal model fit gives an estimate of $R_{25}$ for each galaxy (the radius 
at which the profile intercepts 25 mag/arcsec$^2$ in $F475W$). The foreground galaxy's $R_{25}$ occurs at 80$\pm$1 
pixels or 4\farcs0 (4.5 kpc at z=0.06).

\section{\label{s:map}Optical Depth Maps}

To compare the different methods of deriving extinction, the optical depth maps for the overlap section are shown in 
Figure \ref{f:map} for method A, B, and C and filters $F475W$, $F606W$, and $F814W$).
The full optical depth map from Method C is shown in Figure \ref{f:mapC} for comparison. 

The optical depth maps in Figures \ref{f:map} have numerous features. Most noticeable is the clear presence of 
spiral arms. These features suggest that a cold, dense ISM does exist at larger radii, and that 
it remains concentrated in the spiral arms.
Figure \ref{f:map} shows little extinction in between the spiral features.
These features are most prominent in $F475W$ due to the increase in dust absorption at shorter wavelengths. 
In addition, the effect of substructure in the foreground disk is also the most profound in this filter.
The $F606W$ and $F814W$ filters also show the dust structures, but at progressively smaller optical depths. 

From Figure \ref{f:map} it is clear that the automated fit approach has a significant drawback for this pair. Because 
of the many foreground stars associated with NGC 253 and the resolved structure in the foreground galaxy, using rotated 
images leads to artifacts in the resulting optical depth map. The other two approaches suffer much less from this effect. 
Because the bulge of the background galaxy is largely axisymmetric, method B and C generally agree well, and the dark 
extended structures can be identified in both. Notably, the nearly opaque spiral arm can be seen in all three types of
optical depth map, independent of the method used.

\subsection{Distribution of Extinction}

To quantify the distribution of extinction seen in Figure \ref{f:map}, we plot in Figure \ref{f:hist} histograms of the optical depth in each 
pixel for all three filters and all three methods. 
%
The distribution of optical depths shows the broadest distribution of optical depth values in the $F475W$ band and 
the narrowest in the $F814W$. It also shows that the median of the distributions shifts to lower values with redder 
filters. This effect is expected because the bluest band is the most affected by extinction.
In each filter, the optical depth stays below one, which is consistent with the paucity of molecular gas seen at large radii in nearby spirals. 

Method C  shows no peak, and instead, the distribution of extinction rises smoothly towards zero. 
In contrast, for methods A and B, the distribution of optical depths peaks at $\tau$ = 0.25 for $F475W$ and $F606W$ and 
$\tau$ = 0.15 for the $F814W$ band with  Method A having the most extreme extinction values due to the additional structure in the subtracted foreground galaxy. 

This shift to higher optical depths can be explained by the construction of the background light model. Because methods A and B use the background galaxy itself, any dust structures in the background galaxy {\em increase} the inferred 
extinction distribution in the foreground galaxy. Method C is not affected by this contaminant due to the use of an isophotal light model. 

\subsubsection{Zeropoint Check }

We can verify that Method C's extinction values are unaffected by the contamination of substructure by applying the same method to two apertures where we expect the mean extinction to be close to zero. Figure \ref{f:check} shows the resulting distribution of optical depth values using Method C for three apertures, one in the overlap region, and two in the mostly unobscured region (Figure \ref{f:mapC}). In both low-extinction apertures there is a clear and narrow peak at $\tau=0$. Similar plots for Method A and B also show a peak at $\tau=0$, however with a wider distribution of values around zero. This wider spread of optical depth values around the zeropoint can be attributed to asymmetric structure in the background galaxy. 

Unlike for the unocculted regions, the distribution of extinctions for the occulted region is clearly skewed to significantly higher values. Comparing to the unobscured regions suggests that these higher extinction values in the overlap region are likely to be physical rather than an artifact of the background  galaxy substructure.

\subsubsection{Uncertainty Estimate}

To obtain a simple estimate the uncertainty in the optical depth values in Method C, we shift the model background and foreground galaxy by a pixel in both x and y direction (x+1, y+1 and alternatively x-1 and y-1). Because asymmetry is one of the dominant sources of uncertainty in the occulting galaxy extinction measure, a shift of the center of these models is a straightforward way to estimate the uncertainty in method C's optical depth measurement. The fit of the center of these galaxies is unlikely to be off by more than a pixel. Figure \ref{f:pertChist} shows the differences between the optical depth map values in the overlap region in Figure \ref{f:map}. The change in optical depth as a result of a pixel shift is less than 0.05 magnitude. The uncertainty is less than the difference with Method A or B Figure \ref{f:check})  and it is likely of similar order as the uncertainty due to asymmetry in the general structure of both galaxies (see Figure \ref{f:isomod}).

\subsubsection{Exponential Fit to the Distribution of Extinctions}

The distribution of extinction values is an essential prior to the Bayesian fits of SN1a lightcurve fits 
\citep[see discussions in][their Appendix A]{Wood-Vasey07,Jha07}. The Bayesian approach with an extinction 
prior was first used by \cite{Riess96a} and in much subsequent work. However, the well-known danger of using the 
extinction prior is that it may be in error and, thus, propagate systematic errors into the final distance estimate. 
The models by \cite{Hatano98}, \cite{Commins04}, and \cite{Riello05} show that for late-type galaxies, the likelihood distribution 
for extinguished lines of sight follows an exponential function with a maximum at zero extinction: 
\begin{equation}
\label{eq:dist}
n = n_0 e^{-\tau/\tau_0}.
\end{equation}
The common exponential scale used is $\tau_0$= 0.5 in Johnson $V$.  \cite{Jha07} use a series of Monte-Carlo 
simulations to characterize the effect of a wrong prior and conclude it can result in a significant bias in the 
distance scale. 

In Figure \ref{f:histC}, we show the distribution of extinction values for Method C, along with exponential fits to the positive values. 
The $\tau_0$ values decrease with wavelength as expected: $\tau_0$ is 0.28, 0.15, and 0.09 for $F475W$, $F606W$, and 
$F814W$, respectively. These decay rates are much smaller than the value of 0.5 commonly used as the SNIa prior.
Moreover, the distributions do not continue to rise as an exponential all the way to no extinction. The peak in distribution at zero extinction is smeared somewhat by measurement uncertainties. However, assuming that equation \ref{eq:dist} holds for all values of $\tau$ would lead one to overestimate the fraction of the disk that has low extinction values.

The caveats are that the distribution we measure, is not necessarily the same that the one SNIa experience. Our distribution is for the outskirts of a 
smaller late-type spiral and includes extinction by dust through the entirety of the disk's height, rather than for what one might expect for 
an embedded source. 
On the other hand, our area weighted measurement may be representative for a smooth distribution of sources, such as one might expect for an old population of SNIa precursors. An additional issue is the $z$-distributions of dust and the target objects. The three-dimensional distribution of SN Ia then factors into how the extinction statistics should be used to form a prior. We hope to obtain a large sample of optical depth distributions in occulting pairs with HST data to further specify these distributions and their variance among spirals.

\subsection{Radial Extent of the Dust Extinction}

\begin{figure}[h]
\begin{center}
  \includegraphics[width=0.8\textwidth]{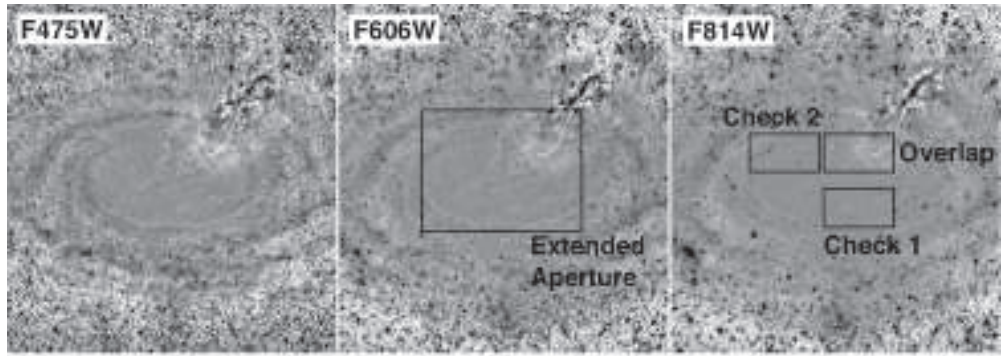}
\caption{The optical depth maps in $F475W$, $F606W$, and $F814W$ from Method C for the pair. A nearly opaque spiral structure is visible extending out of the foreground spiral's disk. Near the center of the background galaxy, a second spiral arm is visible. The spiral pattern of the background galaxy is also visible in each map. The black rectangle in $F606W$ is the extended aperture used to construct the radial profile in Figure \ref{f:ra}. The apertures in the F814W map are the overlap region used for Figures \ref{f:hist}, \ref{f:histC} and \ref{f:ra}  and the check-apertures in Figure \ref{f:check} to verify the zeropoint.  }
\label{f:mapC}
\end{center}
\end{figure}

\begin{figure}[h]
\begin{center}
\includegraphics[width=0.8\textwidth]{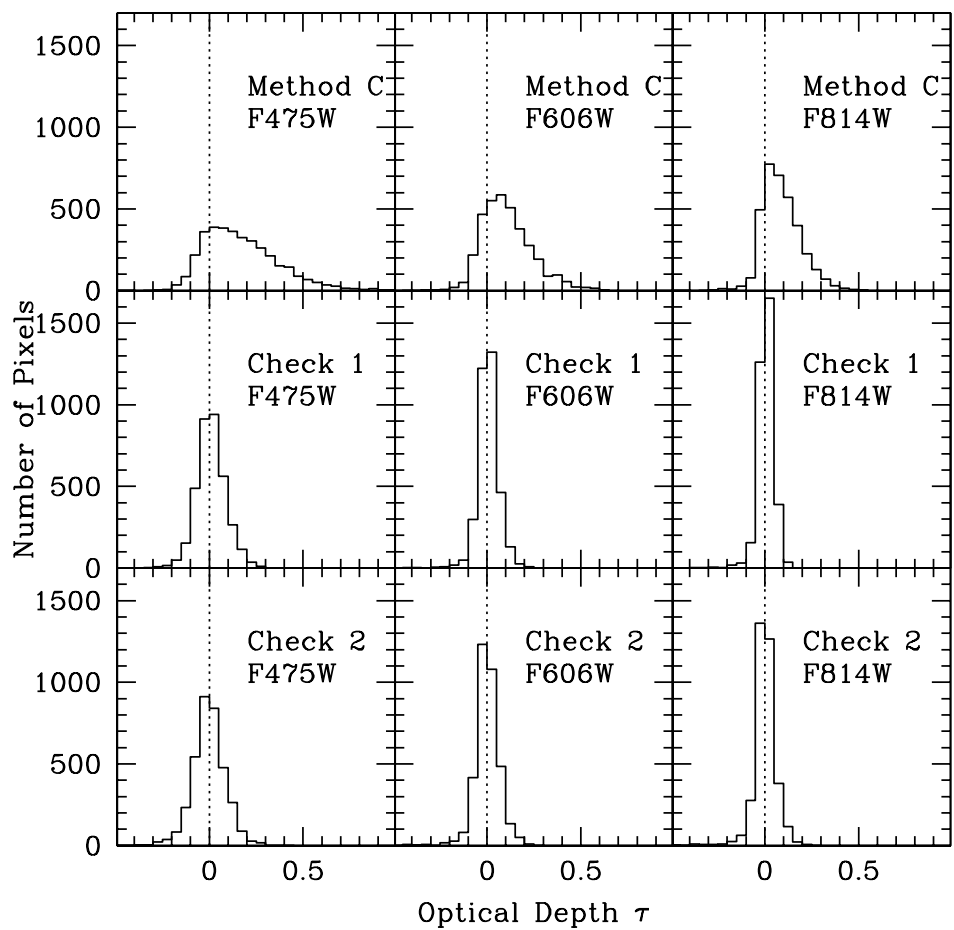}
\caption{The optical depth histograms for $F475W$, $F606W$ and $F814W$ from Method C in the overlap aperture and two equal size apertures on the bulge of the background galaxy. Check apertures 1 and 2 (see Figure \ref{f:mapC}) are not completely devoid of the extinction signature of the foreground galaxy but show a clear peak at $\tau=0$ in each band. }
\label{f:check}
\end{center}
\end{figure}

The most remarkable feature of this overlapping pair is the extraordinary radial extent of dust in the
foreground spiral. Using the extinction maps in Figure \ref{f:map}, we examine the distribution of extinction values 
as a function of the projected radius from the center of the foreground galaxy. Figure \ref{f:ra} shows dust opacity vs. radius for the overlap region shown in Figure \ref{f:map}, and for the larger 
range in radii both derived with Method C (Figure \ref{f:mapC}). Starting at $\sim$ 3 disk half light radii, where we have good statistical sampling, there is a reasonably steady decline of extinction until $\sim$6 half light radii. 

%
The peak at four half-light radii 
($R ~ = ~ 4 ~ R_{50}$) in all three filters can be attributed to the spiral arm visible in the optical depth 
maps as a single dark feature in Figure \ref{f:map}, which crosses the azimuth of the brightest background 
light and is thus detected with high S/N. The signal of two secondary spiral arms can be seen at 2.5 and 5.5 
$R_{50}$. At small radii, the extinction appears to decline. However, at these radii, we have few pixels and the most significant contribution from light emitted by the foreground spiral and the least flux by the background galaxy.

Figure \ref{f:ra} shows the radial plot that can reliably be derived over the entire background bulge using Method C (Figure \ref{f:mapC}). Because we use a model light distribution for the background galaxy flux estimate rather 
than rotating the galaxy, the extinction profile can be extended to the far side of the center of the background galaxy. 
Figure \ref{f:ra} averages over a large section, and thus the mean value (dashed white line) emphasizes the $R \sim 4 ~ R_{50}$ extinction feature but averages out any smaller dust structures.
The radial profile is similar to other radial extinction plots, for example those derived from the UV/FIR flux ratio by \cite{Boissier04} or counts of distant galaxies in \cite{Holwerda05b}. However, this is the first on to be derived for a single galaxy using this technique. 

Figure \ref{f:raav} shows the average radial profiles for overlap region in Figure \ref{f:map} derived using all three methods. The differences in opacity at smaller radii between the three methods is due to the small coverage of the overlap region at $\sim 2 ~ R_{50}$ and the dominance of foreground galaxy structure at these radii in the case of method A.  
Method B and C are generally in good agreement because the background bulge is reasonably symmetric. Structures such as the the extinction in the foreground galaxy's spiral arms show up well with either method. 
Near the center of the foreground galaxy, Method B is likely less accurate because the assumption that the flux from the background galaxy dominates (B$>>$F) and the use of equation \ref{eq2} is less valid. 
From Figure \ref{f:raav} we can conclude that all three methods agree well at the intermediate radii, but that closer to the center of the foreground galaxy, method A suffers from sampling effects and foreground structure and Method B from the less dominant background flux. Hence, Method C has the widest range of places it is applicable in the foreground disk.

\subsection{Observed Extinction Law}

The extinction law can be characterized by its slope $R$ as a function of wavelength. For a given reddening E(B-V) 
and extinction $A_V$ (= 1.086 $\times \tau_V$), $R$ is expressed as:

\begin{equation}
\label{eq:law}
R \equiv {A_V \over A_B - A_V} = {A_V \over E(B-V)}. 
\end{equation}

This relation can be generalized to other wavelengths and bandpasses, such as those on HST.
We derive the values of $R$ from the model given in \cite{CCM} for reference in Table \ref{t:R}. 
Figure \ref{f:extlaw} plots the values of $A_{F475W}$ and $A_{F814W}$ vs. $A_{F606W}$ for the overlap region in Figure \ref{f:map} 
for Method C. There is a clear relation between the extinction values in the three filters, similar to the 
Milky Way extinction relation from \cite{CCM}. The bootstrap mean and standard deviation of the $A_{F475W}/A_{F606W}$ 
and $A_{F606W}/A_{F814W}$ values for all three methods are in Table \ref{t:R}. The inferred $R$ values are also listed.


One expects the observed relation to be grayer than the intrinsic grain properties would give, due to sampling 
effects at greater distance. If the typical pixel size covers large physical scales, lines of sight with different extinction values are 
combined into a single measurement. The lines of sight with lower extinction dominate the emerging light, biasing the color measurement towards less reddened values. This effect is increasingly effective at short wavelengths, and thus the observed extinction law turns gray.
For the system studied here, each pixel covers an area of less than 57 $\times$ 57 pc, assuming the foreground galaxy is at the 
redshift of the background galaxy. \cite{kw01a} report that the extinction law is grayer when sampled over 
physical scales greater than approximately 60 pc. Our consistency with the Milky Way extinction law in this pair is consistent with this conclusion.

 In Figure \ref{f:extlaw}, we use Method C extinction map to explore the relation between the reddening-extinction relation and sampling, given that Method C retrieves a reasonably well-defined extinction law. We average the model 
and data over 2$\times$2 and 4$\times$4 pixels (0\farcs1 and 0\farcs2 respectively). 
The inferred $R$ values for the averaged images are also listed in Table \ref{t:R}. There is a trend towards a grayer extinction law, as expected.
From the inferred relation between extinction and filter, we can conclude that one is likely to 
see deviations from the Milky Way extinction law in external galaxies due solely to the effects of sampling. The critical 
point is when the extinction and color are measured averaged over a physical scale greater than 60 pc.

\section{\label{s:dis}Discussion}

This serendipitously discovered pair of occulting galaxies is in a nearly perfect geometry for the analysis of dust extinction.
The other advantages are high-quality HST observations in three filters with long integration times. 
This particular pair is therefore ideal for constraining dust properties at larger radii as well as for testing three different approaches for analyzing occulting pairs.

The radial extent of the dust in the foreground disk is the most striking feature of the pair. In a part of the disk 
where there is barely any light from the foreground galaxy, there is significant structured extinction evident in the spiral arms, readily visible in the color image (Figure \ref{f:rgb}). At the corresponding radius on the other side of the disk, even these long exposures do not detect starlight, indicating that the cold dense ISM extends well past the optical limit of the disk. The presence of significant amounts of dust at these radii also suggests that substantial quantities of metals have been transported well outside the starforming regions, either through radial mixing or a galactic fountain.

If this is the typical size of a spiral's dusty ISM, it has implications for studies involving lines-of-sight to the 
distant universe \citep[e.g.,][]{Trenti06, Robaina07} or computed SN1a rates \citep[e.g.,][]{Hatano98,Cappellaro99, 
Goobar02, Riello05, Mannucci07}. 
The typical optical depth at $R_{25}$ is also critical for the Tully-Fisher relation in diameter-limited samples. 
From the inclination effects on large samples, it is commonly inferred that spiral disks are optically thin 
($\tau << 0.1$) at this radius, but it depends greatly whether a spiral arm intersects $R_{25}$, as it does in our foreground galaxy. 

Spiral arms are easily identifiable in the optical depth maps in Figure \ref{f:map} and in the radial profiles in 
Figure \ref{f:ra} and \ref{f:ra}. The width of the arms in the optical depth maps appears to be of the order of 
0.5 $R_{50}$ (0.6 kpc), with three arms showing in the radial plots between 2$\times$ and 6$\times$ $R_{50}$.
The typical value of opacity in the disk depends on the filter; the bluest filters, $F475W$ and $F606W$, have a 
typical optical depths of 0.25, while in $F814W$ it is 0.15 (Figure \ref{f:hist}). The range in optical depth values 
becomes much larger in the spiral arms.

The extinction distribution in a spiral disk has wide implications for other observations of spiral galaxies, both 
nearby and at higher redshift. We find different distributions for our different methods, but Method C seems the 
best constrained, and the extinction distribution rises smoothly to $\tau=0$ for Method C. In the case of Method C, 
the distribution can be well modeled with an exponential. Compared with the distribution of values commonly 
used for SNIa light curve fits, these fall off more steeply with increasing $\tau$. 
More occulting pairs would be needed to 
accurately describe the distribution in just the spiral arms and over a range of Hubble types and redshifts. 
This distribution could then be used to form priors for the SNIa lightcurve fits.

\subsection{Best Method}

The best approach for analyzing an occulting pair depends very much on the type of pair and the quality of the data.
For poorly resolved data, Methods A and B are more attractive because their drawbacks --the effects of substructure-- are 
less pronounced. The main attractions of Methods A and B are the speed at which they can be applied. 
In the case of higher resolution data, such as these HST images, Methods B and C make more sense because there is enough 
information in the image to derive results. Method B does rely on the assumption that there is little flux from the 
foreground galaxy, which in this case only holds true at larger projected radii away from the foreground galaxy center.
%
Method C allows us to estimate optical depth to higher radii and estimate the contribution by the foreground 
galaxy closer to the foreground's center. When the data quality permits, this technique is preferred.

None of the approaches we tested can be completely automated at present. All approaches require the position of an aperture 
in which to measure the optical depth in a poorly resolved pair. Method B's assumption needs to be validated and 
Method C requirers visual masking of sections affected by extinction or structure.
We mostly automated Method A for \cite{Holwerda07c}, and an new automated version of Method C could be applied to 
occulting pairs of various galaxy types found by the GalaxyZOO project or at higher redshift HST surveys such as DEEP2 and COSMOS.
 In this case, a S\'ersic model or isophotal profile of both galaxies would be part of the fit solution to the image of the pair.

\section{\label{s:con}Conclusions and Future Work}

\begin{figure}[h]
\begin{center}
  \includegraphics[width=0.8\textwidth]{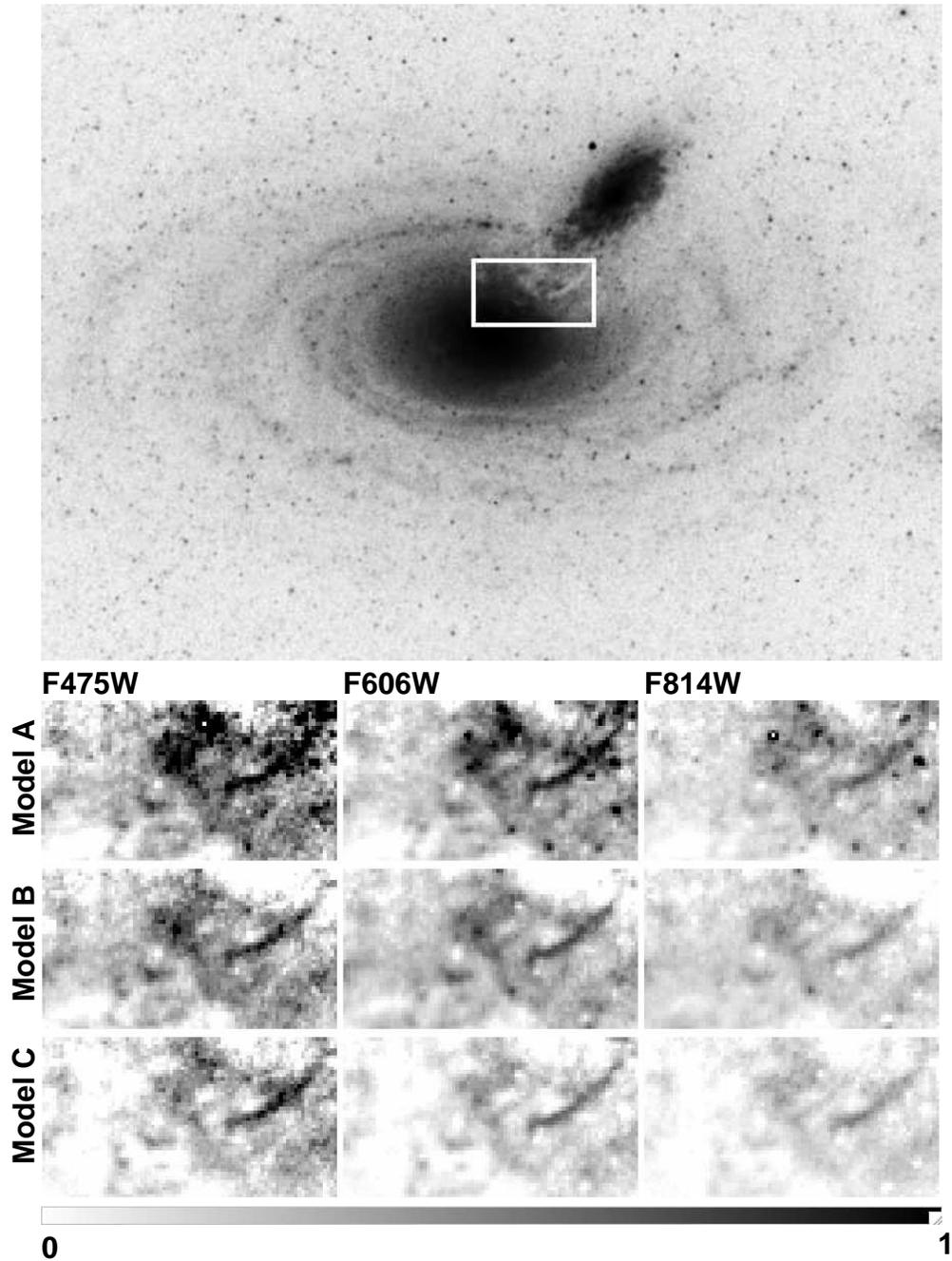}
\caption{The optical depth maps in $F475W$, $F606W$, and $F814W$ for the overlap region, indicated in the $F475W$-band negative image. The scale of the optical depth maps is matched to range from $\tau$=0 to 1. Each row of optical depth maps has been derived with the three different approaches (Method A, B, and C).}
\label{f:map}
\end{center}
\end{figure}

\begin{figure}[h]
\begin{center}
  \includegraphics[width=0.8\textwidth]{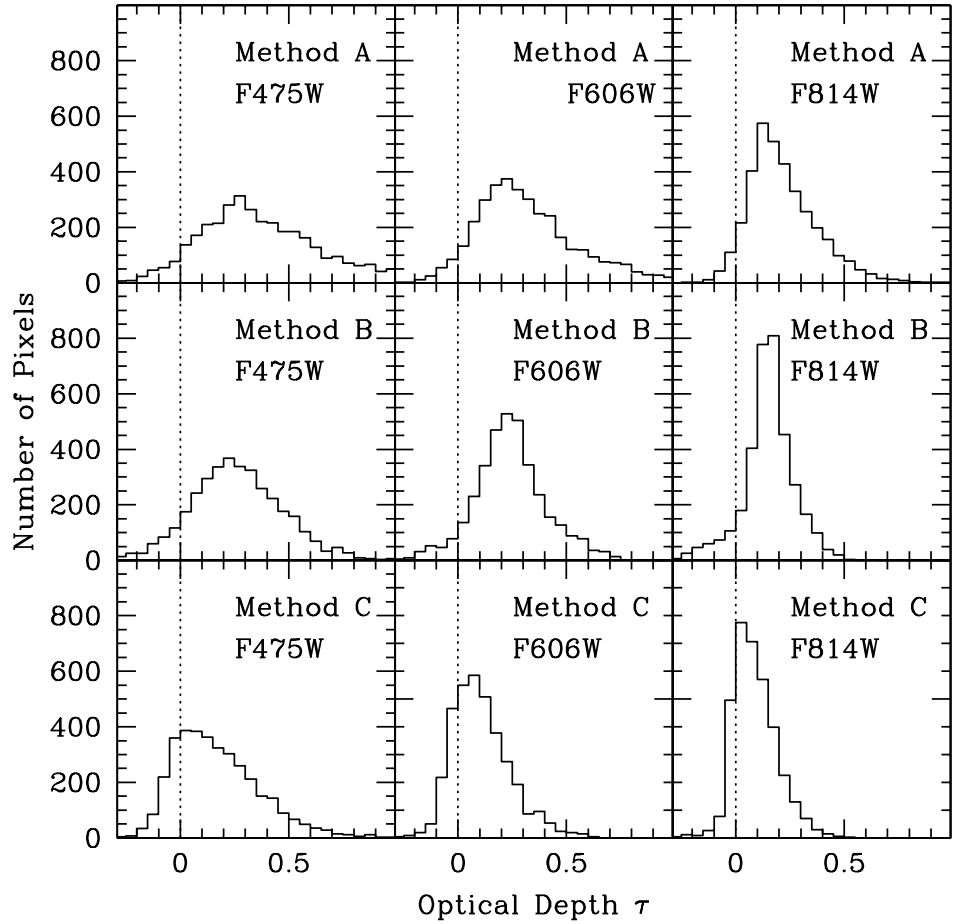}
\caption{The distribution of $\tau_{F475W}$, $\tau_{F606W}$, and $\tau_{F814W}$ in the overlap region indicated in Figure \ref{f:map} for each of the approaches, Method A, B, and C.}
\label{f:hist}
\end{center}
\end{figure}

\begin{figure}[h]
\begin{center}
\includegraphics[width=0.8\textwidth]{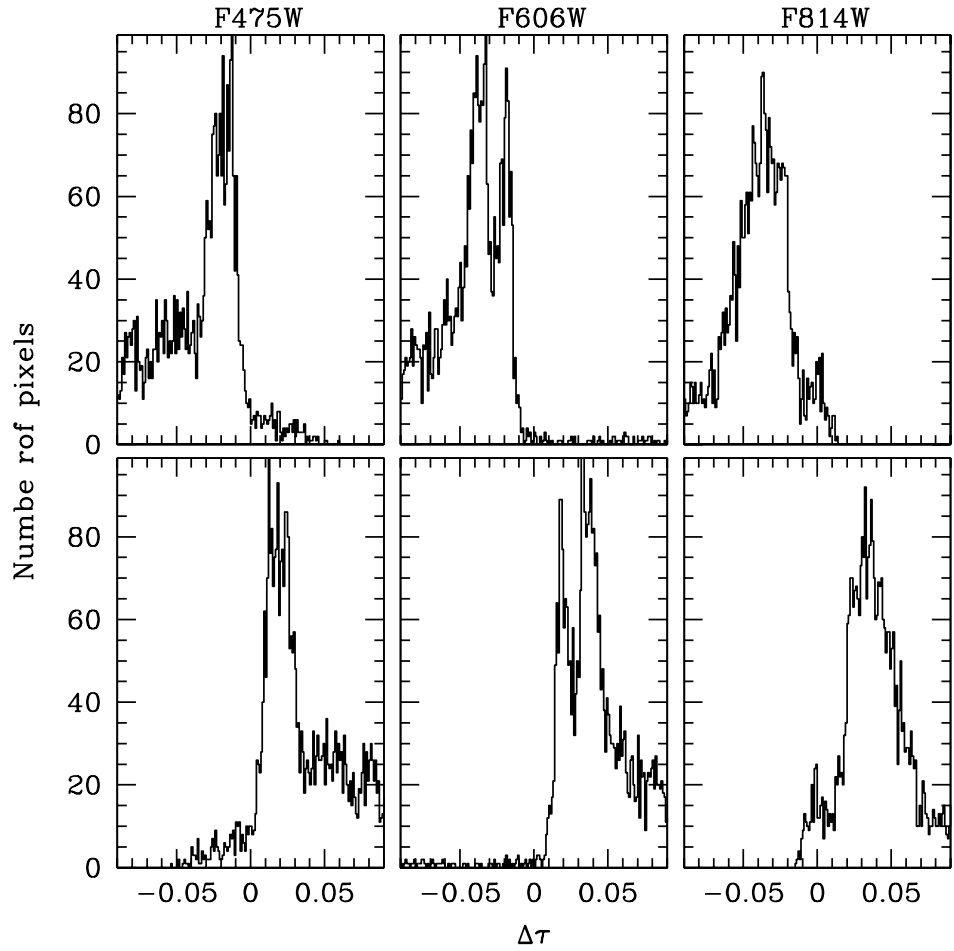}
\caption{The differences between optical depth from Method C with those determined with shifted foreground and background galaxy models for $F475W$, $F606W$ and $F814W$ (Top panels; x+1, y+1, bottom panels; x-1, y-1). 
The differences are smaller than the effects of structure in Method A (Figure \ref{f:hist}). }
\label{f:pertChist}
\end{center}
\end{figure}

\begin{figure}[h]
\begin{center}
\includegraphics[width=0.8\textwidth]{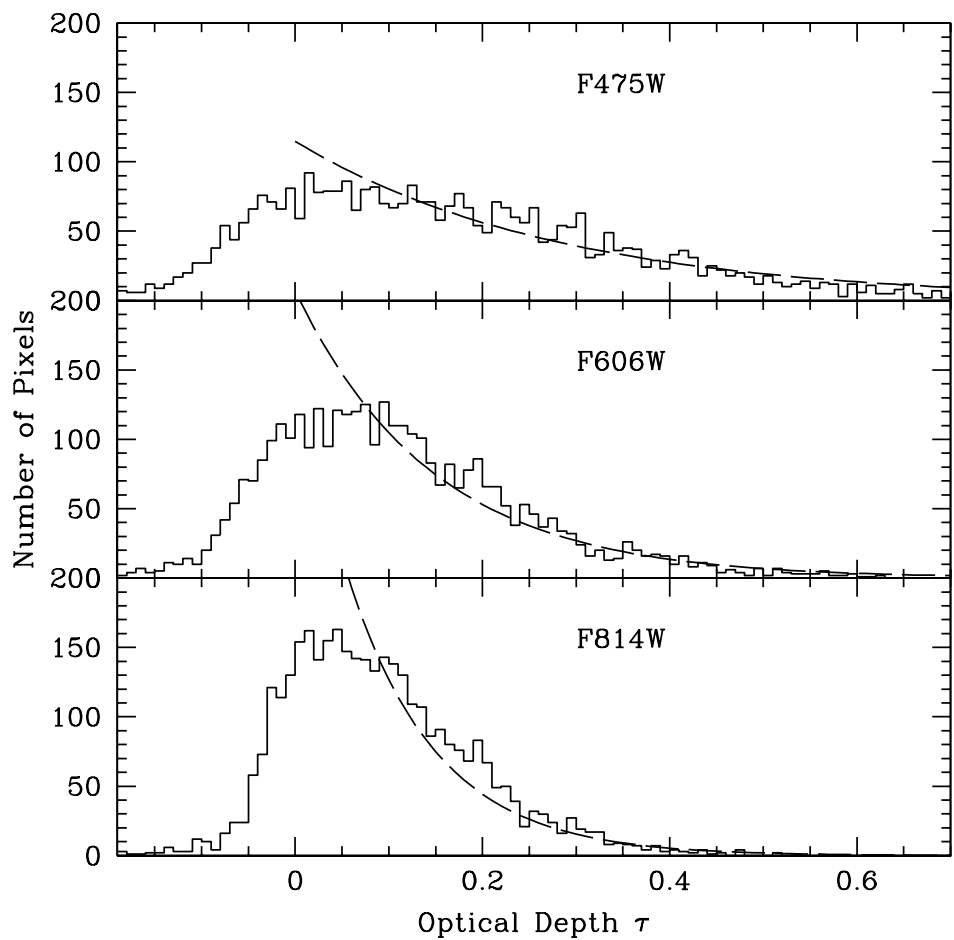}
\caption{The optical depth histograms for $F475W$, $F606W$ and $F814W$ from model C in the overlap aperture (see Figure \ref{f:mapC}) with a finer resolution than Figures \ref{f:hist} and \ref{f:check}. 
We fit $n = n_0 e^{-\tau/\tau_0}$ to the distribution (dashed line) for $\tau > 0.1$. The values of $\tau_0$ are 0.28, 0.15 and 0.09 for $F475W$, $F606W$ and $F814W$, respectively.}
\label{f:histC}
\end{center}
\end{figure}

\begin{figure}[h]
\begin{center}
  \includegraphics[width=0.8\textwidth]{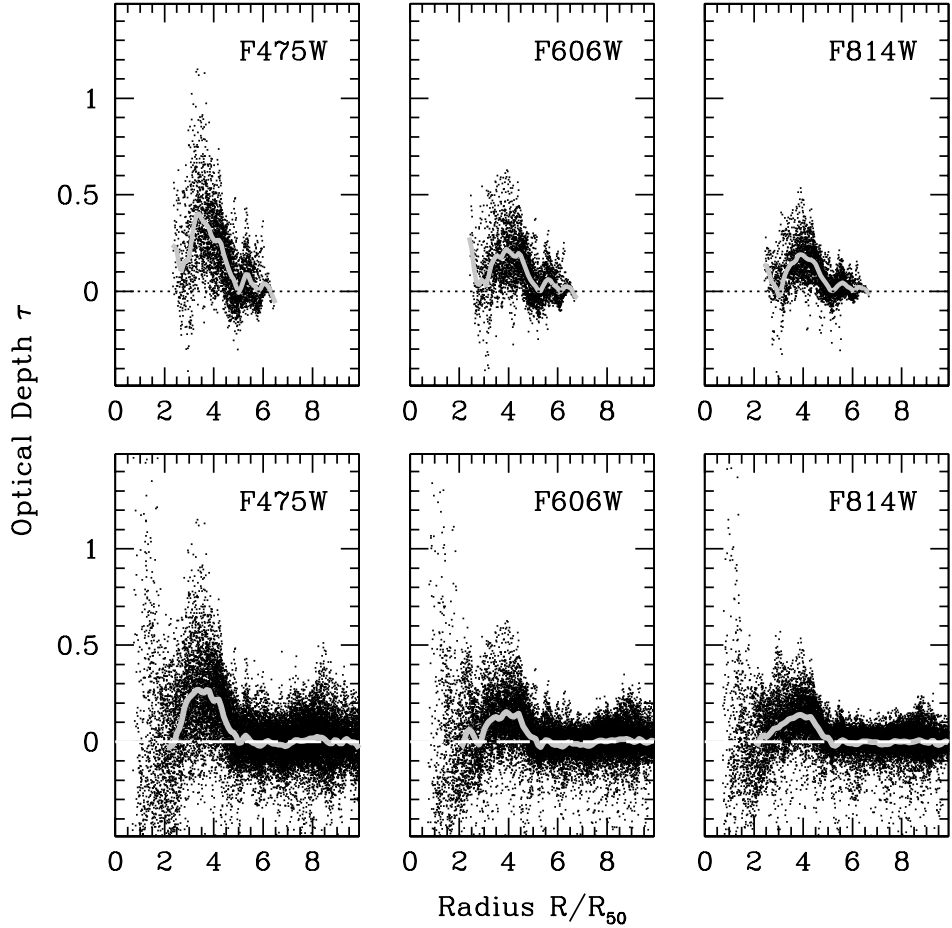}
\caption{{\bf Top}: the radial distribution of $\tau_{F475W}$, $\tau_{F606W}$, and $\tau_{F814W}$ in the optical depth map of the overlap region using method C (Figure \ref{f:map}). The radius is expressed as the effective radius ($R_{50}$, the radius containing 50\% of the light) of the foreground galaxy. The thick gray lines are the mean value of the radial profile. {\bf Bottom}: The radial distribution of optical depth for the three different filters in the extended overlap region based on the optical depth map from Method C. The profile extends to beyond the position of the background galaxy's center. Due to the large area with no extinction in this aperture it nearly averages out to zero but the combined effect of the spiral arms at R = 3 and 4 $R_{50}$ can be seen.}
\label{f:ra}
\end{center}
\end{figure}

\begin{figure}[h]
\begin{center}
  \includegraphics[width=0.8\textwidth]{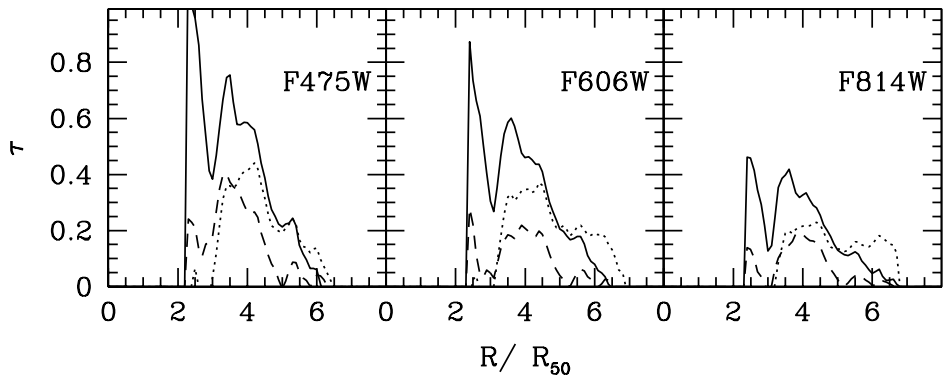}
\caption{The average radial distribution of optical depth for the three different filters and the three different approaches, Method A (solid), B (dotted) and C (dashed, same as the gray lines in Figure \ref{f:map}).  The radius is expressed the effective radius ($R_{50}$) of the foreground galaxy. Method A measures the highest optical depth profile, Method B has trouble near the center of the foreground galaxy, and Method C measures lower values than A or B.}
\label{f:raav}
\end{center}
\end{figure}

\begin{figure}[h]
\begin{center}
  \includegraphics[angle=-90, width=0.8\textwidth]{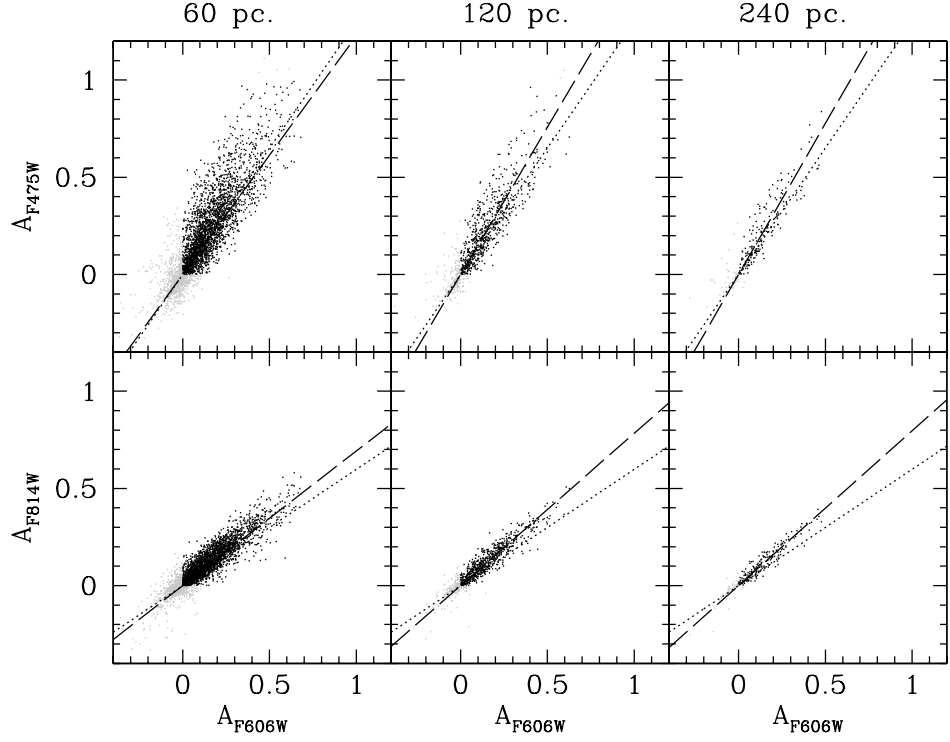}
  \caption{The relation between $A_{F606W}$ vs. $A_{F475W}$ and $A_{F814W}$ for Method C and derived from binned images (2 and 4 pixels, 120 and 240 pc respectively). The black dotted lines are the Milky Way extinction law relation for the ACS filters calculated from \cite{CCM}. The dashed lines are bootstrap mean ratios of the optically thin points (black). Negative or optically thick measurements are marked with gray points. The bootstrap mean values are in Table \ref{t:R}. There is a trend to lower values of R (a grayer relation) with greater sampling size. }
\label{f:extlaw}
\end{center}
\end{figure}

From this occulting pair with a nearly ideal geometry, serendipitously imaged by HST, we can learn the following:

\begin{itemize}

\item[1.] Most of the disk is optically thin ($\tau < 1$) in $F475W$, $F606W$ and $F814W$, from 3 $R_{50}$ to 10 $R_{50}$ (Figures \ref{f:hist} and \ref{f:ra}). 
\item[2.] The distribution of optical depth values follows an exponential with scales of 0.28, 0.15, and 0.09 for $F475W$, $F606W$, and $F814W$ respectively (Figure \ref{f:histC}). At lower optical values close to zero, the exponential distribution overpredicts this distribution some.
\item[3.] Measurable extinction extends to six half-light radii or well beyond the optical disk (1.5 $R_{25}$) of the foreground spiral (Figure \ref{f:ra} and \ref{f:ra}).
\item[4.] Spiral structure is visible as regions of high extinction beyond $R_{25}$ (Figures \ref{f:map} and \ref{f:ra}), confirming that there is a dense cold ISM confined to spiral arms, even at large radii.
\item[5.] The slope of the observed extinction law is similar to the Milky Way extinction law (Figure \ref{f:extlaw} and Table \ref{t:R}).
\item[6.] Averaged over physical scales greater than 60 pc., the observed extinction law is grayer, because the observed light is biased towards the low-extinction lines-of-sight (Figure \ref{f:extlaw} and Table \ref{t:R}).
\item[7.] All three approaches for measuring optical depth have their specific uses, and either Method A or C could be used effectively for future automated analysis of large samples of occulting pairs.
\item [8.] If the data is of high enough quality, Method C is to be preferred, because it introduces less noise. Asymmetry and substructure are the dominant sources of uncertainty in the other two methods (section 5.1)
\item[9.] Due to the likely $30^\circ$ inclination of the foreground spiral, the optical depths we measured are likely to be 15\% higher than for a perfectly face-on geometry.

\end{itemize}


For future studies, one would need a very large sample, such as the 800 occulting pairs identified by 
the Galaxy Zoo project \citep{galaxyzooAAS,galaxyzoo} in the Sloan Digital Sky Survey, or the many occulting pairs 
in high-redshift HST imaging surveys, such the Extended Groth Strip and COSMOS \citep{Koekemoer07}. With such a large sample, the effect of inclination of the foreground disk on the observed optical depth and the evolution of optical depth since 
a redshift of one can be determined \citep{Holwerda07aas,Holwerda08op}.

\begin{figure}[h]
\begin{center}
  \includegraphics[width=0.8\textwidth]{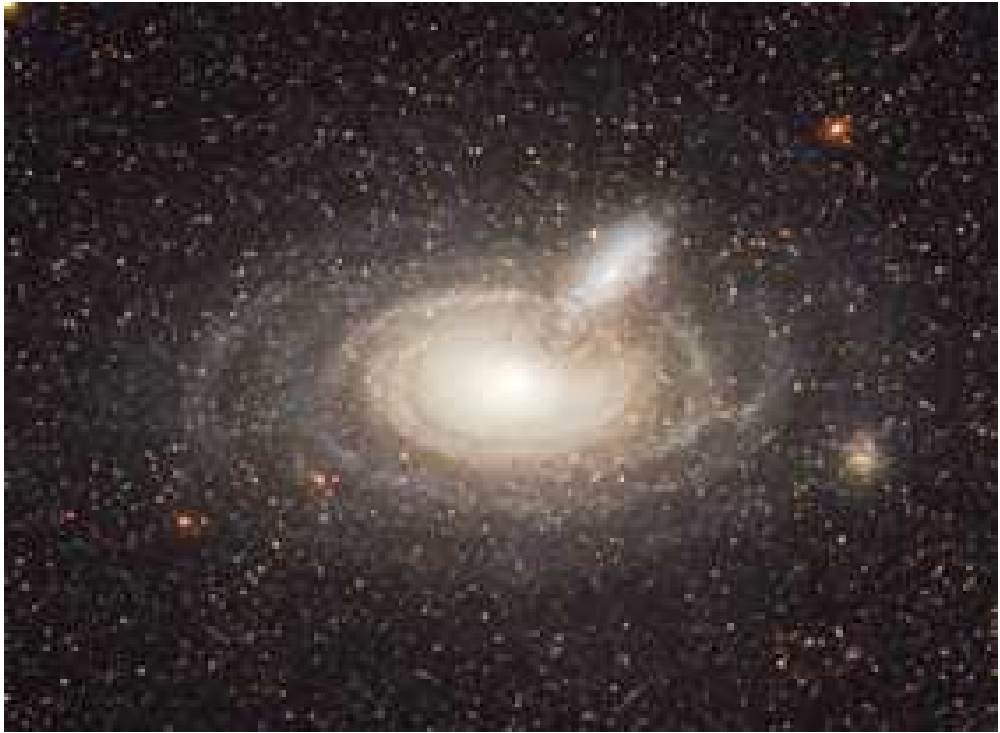}
\caption{The RGB color composite of the pair with $F475W$ for blue, $F606W$ for green, and $F814W$ for red. The RGB stars of the halo and disk of NGC 253 contaminate the foreground. The effect of the foreground galaxy's dust can be seen extending to the center of the central background galaxy. Image thanks to Zolt Levay of the Hubble Heritage Team. }
\label{f:rgb}
\end{center}
\end{figure}

The authors would like to thank Ron Allen for useful discussions and Kristen Keener for her comments and edits. The authors would like to thank Zolt Levay for Figure \ref{f:rgb}. W. C. Keel acknowledges support from a College Leadership Board faculty fellowship. J.J. Dalcanton and B. Williams were partially supported by a grant from NASA (GO-10915), through the Space Telescope Science Institute, which is operated by the Association of Universities for Research in Astronomy, Inc., under NASA contract NAS 5-26555. J.J. Dalcanton was partially supported by a Wyckoff Faculty Fellowship.
Based on observations made with the NASA/ESA Hubble Space Telescope as part of program GO-10915, P.I. J.J. Dalcanton.




\begin{table}[htdp]
\caption{Basic data on the foreground and background galaxies in V}
\begin{center}
\begin{tabular}{l l l l l l l}
galaxy		& RA			& Dec		& $m_V$	& PA 	& Incl.	& $R_{50}$\\ 
			& deg.		& deg.		& mag	& deg.	& deg.	& pixel\\	
\hline
\hline
foreground	& 12.088910	& -25.126517	& 18.15	& 66.7	& 29.2	& 21.1 \\
background	& 12.090799	& -25.126916	& 16.5	& -48.9	& 30.9	& 32.7 \\
\hline
\end{tabular}
\end{center}
\label{t:bdat}
\end{table}%

\begin{table}[htdp]
\caption{The ratio of $A_{F475W}$ and $A_{F814W}$ over $A_{F606W}$ and the corresponding $R$ values: $R_{F475W-F606W}  = A_{F606W}/(A_{F475W}-A_{F606W})$ and $R_{F606W-F814W}  = A_{F606W}/(A_{F606W}-A_{F814W})$. The equivalent values for \cite{CCM} are given and those for the resampled Method C (Bin 2 and 4).	The bootstrap mean and uncertainty are determined for the optically thin ($A>0, A<1$) values.}
\begin{center}
\begin{tabular}{l l l l l}

\hline
Origin		& $A_{F475W}/A_{F606W}$ & $A_{F475W}/A_{F814W}$ & $R_{F475W-F606W}$ & $R_{F606W-F814W}$ \\
\hline 
CCM			& 1.297641999		& 1.671395795		& 3.36				& 2.49\\
\hline
Method A   	& 1.2$\pm$0.01       & 1.4$\pm$ 0.02	 	& 4.4$\pm$0.24		& 2.3$\pm$0.08\\
Method B    	& 1.3$\pm$0.01       & 1.5$\pm$0.02	 	& 3.4$\pm$0.14		& 2.1$\pm$0.07\\
Method C    	& 1.5$\pm$0.02       & 1.3$\pm$0.01	 	& 1.9$\pm$0.07		& 3.6$\pm$0.17\\
Bin 2 (120 pc) 	   	& 1.5$\pm$0.04       & 1.3$\pm$0.02	 	& 2.0$\pm$0.13		& 3.6$\pm$0.24\\
Bin 4 (240 pc)          	& 1.6$\pm$0.06       & 1.3$\pm$0.03  	& 1.8$\pm$0.20		& 3.9$\pm$0.40\\
\hline
\end{tabular}
\end{center}
\label{t:R}
\end{table}%

\end{document}